\documentclass[a4paper, amsfonts, amssymb, amsmath, reprint, showkeys, nofootinbib, twoside]{article}
\usepackage[english]{babel}
\usepackage[utf8]{inputenc}
\usepackage[colorinlistoftodos, color=green!40, prependcaption]{todonotes}
\usepackage{amsthm}
\usepackage{amsmath}
\usepackage{amssymb}
\usepackage{mathtools}
\usepackage{physics}
\usepackage{xcolor}
\usepackage{multicol}
\usepackage{complexity}
\usepackage{graphicx} 
\usepackage{adjustbox}
\usepackage{placeins}
\usepackage[margin=1in]{geometry}
\usepackage[T1]{fontenc}
\usepackage{csquotes}
\usepackage{float}
\usepackage{authblk}
\usepackage{braket}
\usepackage{algorithm2e}
\SetKwComment{Comment}{/*}{*/}
\RestyleAlgo{ruled}

\begin{document}
\title{A Quantum-Inspired Algorithm for Graph Isomorphism}

\author[1,2]{Innes L. Maxwell}
\author[1]{Stefan N. van den Hoven}
\author[1,2,3]{Jelmer J. Renema}
\affil[1]{\footnotesize{MESA+ Institute for Nanotechnology, University of Twente, P.O. box 217, 7500 AE Enschede, The Netherlands}}
\affil[2]{Department of Applied Physics and Science Education, Eindhoven University of Technology, P.O.  Box 513, 5600 MB Eindhoven, The Netherlands}
\affil[3]{Department of Electrical Engineering, Eindhoven University of Technology, P.O.  Box 513, 5600 MB Eindhoven, The Netherlands}
\date{\today} 

\maketitle

\begin{abstract}

The Noisy Intermediate-Scale Quantum (NISQ) era of technology in which we currently find ourselves is defined by non-universality, susceptibility to errors and noise, and a search for useful applications. While demonstrations of practical quantum advantage remain elusive in this era, it provides space to develop and analyze the advantages and limitations of systems and their ability to solve problems. In this work, we critically assess a proposed quantum algorithm for the graph isomorphism problem, implemented on a photonic quantum device. Inspired by the nature of this quantum algorithm, we formulate a necessary condition for the isomorphism of graphs encoded in Gaussian boson samplers and a classical algorithm to test for it. Our classical algorithm makes use of efficiently computable statistical properties of a quantum sampling system to show a pair of graphs fail to meet our necessary condition and thus cannot be isomorphic. We analyze our algorithm in the context of the inspiring, sampler-based quantum algorithm of Bràdler et. al., the classical color refinement algorithm, and the state-of-the-art quasi-polynomial Babai algorithm. 
\end{abstract}

\begin{multicols}{2}

\section{Introduction}\label{sec:intro}
As Noisy Intermediate-Scale Quantum (NISQ)-era technologies continue to mature, a critical eye must be turned to the potential algorithmic applications of these platforms. Several notable NISQ-era claims of quantum advantage come from sampling systems \cite{USTC, USTC25, GoogleWillow}, but these platforms are not yet mature enough to run an algorithm that might demonstrate functional quantum advantage. Robust methods for problems that classical computers struggle to solve efficiently would demonstrate said advantage, motivating the development of algorithms capable of solving classically hard problems. Shor's quantum algorithm for integer factorization \cite{Shor} inspires further exploration of problems similarly located in the complexity theoretic landscape, such as the subject of this work; graph isomorphism, the classical problem of if two discrete graphs are equivalent up to permutation of their vertex labels \cite{Kbler1993}.\par 

Research on graph isomorphism is split between practical, non-general methods for specific kinds of graphs and theoretical, general algorithms that answer isomorphism for any pair of graphs. While specific efficient methods exist \cite{Kbler1993}, the state-of-the-art classical algorithm published by Babai in 2015 \cite{Babai} remains the most efficient general approach to the problem. Given the method's quasi-polynomial efficiency $(2^{\mathcal{O}(\log V)^3}$ for $V$ vertices \cite{Babai, Helfgot}), the ability to develop a polynomial time, general quantum algorithm for graph isomorphism would be a meaningful demonstration of functional quantum advantage. \par 

Quantum algorithms for graph isomorphism exist and make use of various NISQ-era methods, such as annealing \cite{Gaitan2014, SolvAnneal, RBM, Zick15} and quantum walks \cite{WalksGI}. However, only one algorithm has been developed for a platform recently used in a quantum advantage claim \cite{USTC, USTC25}; the quantum sampling-based algorithm of Brádler et. al. \cite{GIGBS} for a Gaussian Boson Sampler. Gaussian Boson Sampling (GBS) \cite{Kruse} is a scheme where squeezed coherent states are injected into and evolved according to a linear interferometer before samples are collected in the form of click patterns by single-photon detectors at the system's output. Each sampler has an associated probability distribution, determined by the input states and properties of the interferometer, that each of these measured samples is `drawn' from. Can being able to collect many samples from a physical system be used to solve graph isomorphism, and if so, how efficiently? \par 

Gaussian Boson Sampling is a natural platform for solving graph problems as the connectivity and form of a graph can be encoded in the system such that the collected samples reflect it's structure \cite{Encode}. The graphs under study in the isomorphism problem are discrete sets of vertices $V$ and edges $E$, the connectivity of which is described completely by the graph's adjacency matrix. This matrix can be encoded (see Sec. \ref{sec: BSGP}) into the degrees of freedom of a GBS system \cite{Kruse, Encode, Cardin2024}, letting one efficiently collect samples to aggregate information about the graph's structure in an attempt to answer certain questions about it. This encodability has lead to the development of GBS algorithms for a variety of graph problems, such as dense subgraph, max clique, and others \cite{Encode, Arrazola_2018, Oh_2024, Zhu23, Yu2023, Mezher23, Anteneh23}. \par 

In this work, we introduce a necessary condition for the isomorphism between two sampler-encoded graphs, as well as a classical algorithm that tests for it. We analyze the performance and scaling of our algorithm in comparison to the proposed quantum algorithm \cite{GIGBS}, as well as the classical color refinement algorithm \cite{WL}. We discuss the similar computational complexity of these algorithms, as well as their inability to compete with the state-of-the-art method \cite{Babai}.\par  

This paper is organized as follows. Our main claims are collected in Section \ref{sec: claims}, then Section \ref{sec: graphs} introduces the relevant aspects of graph theory and the problem of isomorphism so as to inform the goal of the algorithms discussed throughout. Section \ref{sec: BSGP} discusses the Gaussian Boson Sampling platform, as well as the details of encoding a graph into one for the purpose of solving graph problems. In Section \ref{sec: modes}, we define the statistical values calculated for use in our isomorphism algorithm, as well as discuss their generalization to higher order. We then use these quantities to explicitly define our new necessary condition for sampler-encoded isomorphism in Section \ref{sec: cond}, as well as discuss previous conditions from literature. The algorithmic process and its limitations are discussed in Section \ref{sec: algor}, with pseudo-code provided for the two main processes it uses. Finally, in Section \ref{sec: comp} our algorithm is compared to the inspiring quantum algorithm from \cite{GIGBS}, as well as the classical color refinement algorithm \cite{WL}.

\section{Summary of Claims}\label{sec: claims}
In this section we provide a concise summary of this work's claims and contributions:
\begin{itemize}
    \item We introduce a necessary condition for graph isomorphism based on the mode correlations of graph-encoded Gaussian boson samplers.
    \item We introduce a classical algorithm that distinguishes non-isomorphic graphs using this necessary condition.
    \item We claim our method has similar complexity and graph distinguishing power as the algorithm in \cite{GIGBS}, without the need for a physical quantum device.
    \item We claim that our algorithm, while similar to the classical color refinement algorithm \cite{WL} in complexity, assesses and compares different information about the graphs and is thus a new, distinct method. It is an open question if a result similar to \cite{CFI, CFI2} exists to define cases our method cannot solve.
    \item We claim that at maximum computational order, equivalent to the number of vertices in the graph $G$, our method recovers the relevant probability distributions and is able to sufficiently test for graph isomorphism, but not efficiently.
\end{itemize}

\section{Graph Theory and Isomorphism}\label{sec: graphs}
A graph $G$ is a set of vertices $V$ that are connected in some way by a set of edges $E$. These edges exist between pairs of vertices and can be assigned a variety of properties like weight and direction. In this work we restrict our consideration to the case of simple graphs, in which a pair of vertices can only be connected with a single edge, no vertices can be connected to themselves, and all edges are uniform and undirected. These graphs can thus be described as a collection of binary relations between vertices: either a pair is connected or they aren't. This information is contained completely in the graph's adjacency matrix $A$, which for an $M$-vertex graph is an $M\times M$ matrix with elements $A_{i,j}=\{0,1\}$ indicating if an edge exists between vertices $i$ and $j$. \par 
A variety of computational problems have been defined using graphs, often asking questions about the existence of substructures or comparing the properties of pairs of graphs. A notable problem is graph isomorphism, which asks if two graphs are completely structurally identical, up to the labeling of their vertices \cite{Kbler1993}. More formally, we can describe isomorphism as a relation between graph adjacency matrices.\\
\\
\textbf{Definition 1}: A pair of graphs $G_{1}$ and $G_{2}$ with adjacency matrices $A_{1}$ and $A_{2}$ are isomorphic if and only if there exists some permutation matrix $\sigma$ such that $\sigma A_{1}\sigma ^{T}=A_{2}$ \cite{Kbler1993, Babai, GIGBS}. \\
\\ 
Graph isomorphism is a decision problem, meaning an algorithm should return a yes-no answer when fed two graphs and asked if they are isomorphic. If the returned answer is yes, this can be efficiently verified by applying the resulting permutation to the graphs under consideration. The na\"{i}ve, brute-force approach of manually searching for a permutation between binary $(\{0,1\})$ matrices is computationally hard, necessitating a more clever approach to the problem. As isomorphism between a pair of graphs exists independently of the way they are drawn or labeled, algorithms compare certain immutable properties called graph invariants. A wide variety of invariants exist, describing graph properties like connectivity, edge density, possible colorings, and more. A pair of isomorphic graphs would, by definition, be identical according to any invariant used to compare them \cite{Kbler1993}. However, it is possible for a pair of graphs to be identical according to one or several invariants but not isomorphic. We thus distinguish between invariants which are necessary for isomorphism, and complete invariants, which are sufficient to determine isomorphism. \par 

While a complete invariant allows one to conclusively solve graph isomorphism, none exist that can be implemented efficiently in a test between any two graphs \cite{Kbler1993}. A collection of classical algorithms exist that are capable of solving the problem efficiently for special families of graphs with restricted structures, such as tree and planar graphs. While these invariants are sufficient to distinguish specific families of graphs, they fail to answer the general question of graph isomorphism for any pair of graphs. In state-of-the-art algorithms, such as \cite{Babai}, graphs are often tested according to a collection of necessary invariants to form a complete set.

\section{Boson Sampling and Graph Problems}\label{sec: BSGP}
Boson sampling is a scheme in which $N$ single photons are injected into an $M$-port linear interferometer \cite{AABS}. The $M$-mode interferometer is characterized by an $M\times M$ unitary matrix, the elements $U_{i,j}$ of which correspond to the probability amplitude of a photon traveling from input port $i$ to output port $j$. The photons are collected by single-photon detectors at the output ports of the interferometer, where a pattern of mode occupancy \textbf{n} $=\{n_{1},n_{2},..,n_{M}\}$, in which $n_{i}$ is the number of photons detected in mode $i$, is recorded as a sample. Each output pattern has an associated probability, and the complete collection of the possible output patterns and their probabilities form the probability distribution of the sampler.\par
The task of boson sampling is to draw a sample from the sampler's probability distribution, a process that's been proven hard to simulate exactly using a classical computer, under reasonable assumptions \cite{AABS}. Using a physical sampling system, one can collect samples directly from the encoded distribution of the physical system in order to approximate the values of its elements. Classical simulation attempts to, within some error margin, approximate the probability values of distribution elements. This is done using a variety of techniques, focused around the evaluation of expensive matrix functions. The exact matrix function varies with the type of boson sampling being performed, our choice of which will be elaborated on in the following section. 

\subsection{Gaussian Boson Sampling}
The original Boson Sampling scheme uses single photon (Fock state) inputs \cite{AABS}, which can be difficult to reliably generate in an experimental setting. Replacing the single photon inputs with Gaussian states provides individual-mode tunability via optical pumping power and shifts us to the scheme of Gaussian Boson Sampling (GBS) \cite{Kruse}. Introducing this new degree of freedom enables the faithful representation of a discrete graph using a GBS system, as will be elaborated on in the following section \cite{Encode}. The shift in input state changes the statistics of the system, which is reflected in the probability of observing a specific output pattern \textbf{n} from a Gaussian boson sampler, given \cite{Kruse, COh, AdessoCVQ} by
\begin{equation}\label{eq: probhaf}
p(\textbf{n})=\frac{|\text{Haf}(U_{\textbf{n}})|^2}{\textbf{n}!\sqrt{|\chi + I/2|}}\end{equation}
where $\chi$ is the covariance matrix of the output state, $I$ the identity, and $U_{\textbf{n}}$ the submatrix of $U$ in which rows and columns correspond to the input and output modes occupied by photons, respectively. The covariance matrix elements $\chi_{i,j}$ are two-point correlation values between the modes $i$ and $j$, an idea we will make use of in Sec. \ref{sec: modes}.\par 
The matrix function being applied to the detection event-specific unitary $U_{\textbf{n}}$ is the Hafnian, defined \cite{Kruse} as
\begin{equation}\label{eq: hafnian}
\text{Haf}(A)=\sum_{\rho\in P_{2n}^{2}} \prod_{\{i,j\}\in \rho} A_{i,j}\end{equation}
where $P_{2n}^{2}$ is the complete set of partitions of $2n$ elements into unordered, disjoint pairs. The function sums over the product of elements $A_{i,j}$ of the $2n\times 2n$ matrix $A$ for each of the partitions $\rho$. For example, when $n=2$, the Hafnian yields
\begin{equation}\label{eq: hafex}
\text{Haf}(A)=A_{1,2}A_{3,4}+A_{1,3}A_{2,4}+A_{1,4}A_{2,3}\end{equation}
The Hafnian is classically hard to calculate \cite{Hafnian}, as the size and number of these partitions grows rapidly with the dimension of the matrix. The fastest known  algorithm to exactly compute the Hafnian of an $M\times M$ complex matrix scales as $\mathcal{O}(M^{3}2^{M/2})$ \cite{Hafnian}, rendering it intractable for classical computers at even median values of $M$. One of these calculations must be performed to determine the probability of each possible output, of which there are exponentially many in the number of modes $M$ and photons $N$, making the true probability distribution of a Gaussian boson sampler classically inaccessible through direct calculation. That the probability of each outcome is given by a Hafnian may render exact calculation impossible, but does motivate the applicability of GBS to graph problems, as we will now expound upon.

\subsection{Graph Problems with GBS}
In order to use GBS to study graph problems we must encode the information contained within the adjacency matrix $A$ into the sampling system in order to collect samples that reflect the graph's structure. This is enabled by the Autonne-Takagi decomposition \cite{GIGBS, Encode, COh} for complex symmetric matrices: $A = UDU^{T}$, where the unitary $U$ describes an $M$-mode interferometer and $D$ a diagonal matrix diag$(\lambda_{1},...,\lambda_{M})$ of the $M$-vertex graph's spectrum.  Notably, this condition for decomposition enables weighted and self-looped graphs to be encoded in a GBS device as well. It is also crucial to note that, in order to ensure the resulting unitary matrix is Gaussian, the Takagi factorization is applied to a direct sum of two copies of $A$, $\Tilde{A}=A\oplus A$ \cite{Encode}. The spectrum of this graph is used to determine the input squeezing values, with the squeezing strength $r_{i}$ in mode $i$ defined as $\lambda_{i}=\tanh(r_{i})$. With a corresponding squeezing parameter for each input and an interferometer configured according to the unitary $U$, the sampler's associated probability distribution now contains all of the information in the graph adjacency matrix $A$.\par 

The suitability of a GBS system for solving graph similarity problems follows from the presence of the Hafnian in Eq. (\ref{eq: probhaf}) for a pattern's detection probability. When applied to a graph's adjacency matrix, the Hafnian counts the number of perfect matchings present in said graph. A perfect matching is a subgraph in which every vertex is part of exactly one pair of vertices that share one edge, and the number that can exist grows rapidly with the size of the graph. Each term of the Hafnian's sum is a way to pair up the graph's vertices, and will only be non-zero when each edge between those pairs exists within the original graph. The value of the Hafnian must be equivalent for a pair of isomorphic graphs, as they contain identical substructures \cite{Kbler1993, Babai, GIGBS}. In other words, subgraph structures within pairs of isomorphic graphs must be mappable to one another; each and every submatrix of $A_{1}$ must have a corresponding submatrix in $A_{2}$ for which the Hafnian returns the same value. \par

That the probability of detecting an output pattern $\textbf{n}$ is given by the Hafnian of the transfer matrix $U_{\textbf{n}}$ suggests these detection events reflect the connectivity of the encoded graph. A detection event of $|\textbf{n}|$ photons indicates that one of the possible perfect matchings on as many vertices exists as a subgraph of the encoded graph. The vast quantity of possible perfect matchings on a graph and all of its constituent subgraphs, paired with the exponentially large number of possible detection patterns for a sampling system, makes collecting samples an inefficient way to solve problems, such as isomorphism, by directly comparing samples. However, a method \cite{TwoModeGBS} to calculate lower-order cumulants of the sampler's distributions can be leveraged to more efficiently calculate and compare properties of graph-encoded samplers. The specific conditions for isomorphism when comparing cumulants will be discussed in the following section, along with their graphical interpretation. 

\section{Sampler Mode Correlations}\label{sec: modes}

In \cite{TwoModeGBS}, Phillips et. al. introduce a two-point correlator between pairs of output modes as means to benchmark a Gaussian Boson Sampling experiment. These two-point correlators are the second-order cumulants of two random variables, in this case the number operator evaluated for a pair of modes \cite{TwoModeGBS, Bench}. These two-mode correlations also make up the elements of the covariance matrix $\chi$ from Eq. (\ref{eq: probhaf}), with $\chi_{i,j}$ the correlation between modes $i$ and $j$. \par 
For a pair of classical random variables $X$ and $Y$, the second-order correlation is defined generally \cite{TwoModeGBS} as 
\begin{equation}\label{eq: twocm}
\mathbb{C}(X,Y)=\mathbb{E}(XY)-\mathbb{E}(X)\mathbb{E}(Y)
\end{equation}
In terms of the number operator $\hat{n}_{i}$ in the spatial mode $i$, Eq. (\ref{eq: twocm}) becomes
\begin{equation}\label{eq: twmd}
\mathbb{C}(\hat{n}_{i},\hat{n}_{j})=\braket{\hat{n}_{i}\hat{n}_{j}}-\braket{\hat{n}_{i}}\braket{\hat{n}_{j}}
\end{equation}
The number operators can be written as creation and annihilation operators, $\hat{n}_{i}=\hat{a}^{\dagger}_{a}\hat{a}_{b}$, and decomposed into products of two-operator terms \cite{TwoModeGBS}:
\begin{equation}\label{eq: input}
C_{a,b,c,d}^{(\text{in})}=\delta_{a,c}\delta_{b,d}\braket{\hat{a}^{\dagger}_{a}\hat{a}^{\dagger}_{c}}\braket{\hat{a}_{b}\hat{a}_{d}}+\delta_{a,d}\delta_{b,c}\braket{\hat{a}^{\dagger}_{a}\hat{a}_{d}}\braket{\hat{a}_{b}\hat{a}^{\dagger}_{c}}
\end{equation}
To calculate the correlation values for the output modes of our GBS device, we must evolve the input relation in Eq. (\ref{eq: input}) according to the linear optical unitary: $\hat{a}_{j}\rightarrow \sum_{i}U_{i,j}\hat{a}_{i}$. To do this, we take the product of our mode correlators with the unitary elements that describe paths photons could take to end up in those modes \cite{TwoModeGBS}. Formally, define the two-mode output correlation for an $M$-mode sampler as \cite{TwoModeGBS}:
\begin{equation}\label{eq: cout}
C_{x,y}^{(\text{out})}=\sum_{a,b,c,d=1}^{M}U^{*}_{x,a}U_{x,b}U^{*}_{y,c}U_{y,d} \times C_{a,b,c,d}^{(\text{in})}
\end{equation}
This value represents the summed likelihoods of photons following paths to our output modes of interest from each collection of possible inputs.\par 
Having defined the input mode correlations and the relevant elements of the unitary that determine their evolution, we can then write Eq. (\ref{eq: cout}) as
\begin{equation}\label{eq: cout2}
\begin{aligned}
C_{x,y}^{(\text{out})}=\sum_{a,b=1}^{M}\braket{\hat{n}_{a}}(\braket{\hat{n}_{b}}+1)U^{*}_{x,a}U_{y,a}U_{x,b}U^{*}_{y,b}\\
+\sum_{a,b=1}^{M}\braket{\hat{\epsilon}_{a}}\braket{\hat{\epsilon}_{b}}U^{*}_{x,a}U^{*}_{y,a}U_{x,b}U_{y,b}
\end{aligned}
\end{equation}
where we have replaced the operator terms of the input correlation by quantities determined by each mode's input squeezing parameter, which can be calculated explicitly \cite{TwoModeGBS}:
\begin{equation}
\begin{aligned}
\braket{\hat{a}^{\dagger}_{i}\hat{a}_{i}}    =\braket{\hat{a}_{i}\hat{a}^{\dagger}_{i}}-1=\braket{\hat{n}_{i}}=\sinh^2(r_{i})\\
\braket{\hat{a}^{\dagger}_{i}\hat{a}^{\dagger}_{i}}=\braket{\hat{a}_{i}\hat{a}_{i}}=\braket{\hat{\epsilon}_{i}}=\frac{\sinh(2r_{i})}{2}
\end{aligned}
\end{equation}
where $\hat{n}_{i}$ is again the number operator and $\hat{\epsilon}_{i}$ is the eccentricity of the squeezed state in quantum optical phase space \cite{TwoModeGBS}, both in an input mode $i$. In this work we deviate from the assumption in \cite{TwoModeGBS} that $\braket{\hat{n}_{i}}=\braket{\hat{n}}$, or that the average photon number is the same in each mode. We do not make this assumption, as the value of $\braket{\hat{n}_{i}}$ is determined by the squeezing strength $r_{i}$ in mode $i$, which is in turn determined by the graph eigenvalue $\lambda_{i}$ \cite{Encode}, meaning the input squeezing and thus average photon number per mode will not be uniform. An important consequence of this is that the odd-order correlation terms will not automatically be zero, as they are when squeezing is uniform. The equivalence of graph spectra, or isospectrality, is a frequently implemented necessary condition for isomorphism \cite{Kbler1993}. As the spectra are a pair of $M$-element lists for $M\times M$ graphs, the cost of comparing them is polynomial in $M$. By leaving the graph spectra unmodulated, we are able to efficiently check the isospectrality condition upfront before proceeding with more expensive steps of the algorithm.\par 
Equation (\ref{eq: cout2}) allows us to explicitly calculate the correlation between any pair of output modes in a graph encoded sampler using the input squeezing values $r_{i}$ and unitary elements, which we have full access to from our decomposed graph adjacency matrix.  \par 
In order to calculate correlation between $k$ modes of a sampler, one can use the $k$-order Ursell functions \cite{TwoModeGBS, Bench} for $k$-body correlation. Equation (\ref{eq: twocm}) is the $k=2$ case of the general form \cite{Ursell}:
\begin{equation}\label{eq: ursell}
\begin{aligned}
\mathbb{U}_{k}(x_{1},,x_{k})=&\\
\sum_{Q\in\mathcal{P}_{k}}(-1&)^{|Q|-1}(|Q|-1)!\prod_{Q\in\mathcal{P}_{k}} \left \langle \prod_{q\in Q}x_{q}\right \rangle
\end{aligned}
\end{equation}
where $\mathcal{P}_{k}=\{Q_{1},...,Q_{l}\}_{k}$ is the complete set of $l$ partitions of $k$ elements. Each partition $Q$ contains a unique sorting of the elements into subsets called blocks, denoted $q$. The number operators are grouped into expectation values according to these blocks before being replaced with their corresponding ladder operators, meaning the largest block of the subset is of size $2k$ for $k$ elements. As the output state is Gaussian, any block containing more than two operators can be written as a sum over the complete set of partitions of as many operators into pairs \cite{TwoModeGBS, Bench}. This is equivalent to writing any higher order mode correlation as a sum of products of two-mode correlation terms, which is the method we use to calculate each order of correlation values. Equation (\ref{eq: ursell}) thus describes a linear combination of partitions of $k$ objects with recursively defined prefactors. We can describe the complete set of $k$-order mode correlations $C^{(k)}=\{C^{(1)},C^{(2)},..,C^{(M)}\}$ for the output of a $M$-mode sampler-encoded graph. \par

Linking the values of $k$-order mode correlations directly to a graph property is non-trivial, even in the non-collisional sampling regime. In closed-form expressions developed by Cardin et. al. \cite{Cardin2024}, the terms that contribute to $k$-order cumulants of the sampler are those corresponding to the set of Restricted Perfect Matching Permutations for $2k$-vertex loopless graphs. These matchings are constructed between sets of $k$ vertices, each from one of the copies of $A$ used in the decomposition and encoding. Those that contribute to a $k$-order cumulants have alternating walks between vertices from the graph copies, which excludes terms comprised of multiple lower-order cumulants. The pair-wise collection of operators in terms expanded according to the $k$-order Ursell function, such as eq. (\ref{eq: input}), correspond to connected vertices in this restricted class of subgraphs. A clear visual explanation of the terms that survive this expansion, as well as a generalization to the cases of self-looped graphs and displaced states can be found in \cite{Cardin2024}. Given a graphical description of the specific structures that contribute to the values we will calculate to compare within our algorithm, we will now detail the conditions for sampler-encoded graph isomorphism.\par 

\section{Conditions for Sampler-Encoded Graph Isomorphism}\label{sec: cond}
Once each graph has been encoded in the distribution of a sampler, the question shifts to the means used to test a pair of them for isomorphism. 
\subsection{A Sufficient Condition}
In \cite{GIGBS}, Bràdler et. al. introduce a sufficient condition for GBS-encoded graph isomorphism:\\
\\
\textbf{Definition 2}: A pair of sampler-encoded graphs with adjacency matrices $A_{1}$ and $A_{2}$ and associated encoded probability distributions $P_{1}$ and $P_{2}$ are isomorphic if and only if 
\begin{equation*}
\sum_{\rho \in S_{|\textbf{n}|}}\sqrt{p_{1}(\rho(\textbf{n}))}=\sum_{\pi\in S_{|\textbf{n}|}}\sqrt{p_{2}(\pi(\textbf{n}))} \ \ \ \forall \ \textbf{n}
\end{equation*}
where $\rho(\textbf{n})$ and $\pi(\textbf{n})$ are permutations of the detection pattern \textbf{n} by an element of the symmetric group $S_{|\textbf{n}|}$. The symmetrized sums thus aggregate the output probability for each pattern \textbf{n} under permutation of every member of the symmetric group. The collection of states who's probabilities are summed over for each element $\rho \in S_{|\textbf{n}|}$ forms an orbit, or equivalence class up to permutation. If, hypothetically, one had access to the value of $p$ for each element in the two distributions, it would then be possible to efficiently test a pair of encoded graphs for isomorphism. \\
\\
The isomorphism condition is thus the equivalence of these sums between the samplers for every possible detection pattern \textbf{n}. Meeting this requirement means the distributions and thus encoded graphs are permutationally similar, i.e. they fulfill definition 1 and some permutation matrix $\sigma$ exists such that $\sigma A_{1}\sigma^{\intercal}=A_{2}$. \par 
This sufficient condition means one could use the complete probability distributions of two graph-encoded samplers to conclusively answer the isomorphism question for those graphs. To explicitly evaluate each of the probability terms, of which there a great number, requires calculating a Hafnian \cite{Kruse}, rendering them classically inaccessible \cite{Hafnian}. Even with an ideal physical sampler, it would require exponential time to collect enough samples to approximate the values within polynomial precision. Given the strict equivalence implied in definition 2, we see that neither of the methods used to approximate the distribution of a sampler-encoded graph provides enough information to sufficiently solve graph isomorphism efficiently with a GBS system, regardless of technological progress.

\subsection{Necessary Conditions}
Given the existence of a sufficient condition for sampler-encoded graph isomorphism, the authors of \cite{GIGBS} propose a way to use samples collected from a physical sampler to test for a necessary, but not sufficient, isomorphism condition. Samples are collected from the quantum device and sorted into orbits as a means to coarse-grain the distributions and thus make them easier to compare. They show that the coarse-grained distributions of two isomorphic graphs must be equivalent under permutation, but that the condition is only necessary, not sufficient. While it is insufficient to determine if two graphs are isomorphic, the method does let one conclude that a pair of graphs cannot be isomorphic by showing these coarse distribution terms are not equivalent. Inspired by this, we have formulated a necessary condition to serve as a graph invariant and means to determine non-isomorphism that instead makes use of classically computable statistical properties of the sampler; the aforementioned truncated correlation functions.\par
Under the assumption that a pair of encoded graphs $G_{1}$ and $G_{2}$ are isomorphic, there must exist some permutation $\sigma$ that maps their adjacency matrices to one another. Given the existence of isomorphism, we assert that the correlation values of the sampler modes must also be equivalent up to permutation. Given the definition of correlation functions as sums of $1\leq k\leq M$-order marginal probabilities, we also claim that access to the $k=M$ values would recover the complete probability distribution and enable test of the sufficient condition in definition 2, but at often prohibitive computational expense. \par   
As a pair of identical distributions have identical cumulants, but distributions with identical cumulants are not necessarily identical, this condition cannot be used to sufficiently conclude that a pair of graphs are isomorphic. However, as it is necessary that the cumulants, or correlation functions, be identical, showing that a pair of graphs fail to meet this condition allows one to conclude they are non-isomorphic.\\
\\
\textbf{Theorem 1}: A pair of sampler-encoded graphs with probability distributions $P_{1}$ and $P_{2}$ and set of $k$-order correlations $C^{(k)}_{1}$ and $C^{(k)}_{2}$ are isomorphic if and only if 
\begin{equation*}
\sigma C^{(k)}_{1}\sigma^{\intercal}=C^{(k)}_{2}\ \forall k\in M
\end{equation*}
where $M$ is the number of vertices, and thus maximum order of correlation. \\
\\

Note that the $k$-order correlation values must be permutationally similar under $\sigma$ for all $k$, a property our algorithm takes advantage of, as will be detailed in the following section. 

\section{A Quantum-Inspired Algorithm}\label{sec: algor}

The goal of our algorithm is to compare progressively higher-order sets of mode correlation values for a pair of sampler-encoded graphs to determine if they are isomorphic. More precisely, for a pair of $M$-vertex graphs $G_{1}$ and $G_{2}$ with adjacency matrices $A_{1}$ and $A_{2}$, the goal of our algorithm is to determine if a valid permutation matrix $\sigma$ exists that maps $A_{1}$ to $A_{2}$. As lower-order correlations paint an incomplete picture of the sampler's distribution, showing these values are equivalent does not immediately imply that the encoded graphs are isomorphic. However, as each order of correlation must also be equivalent up to permutation, the algorithm succeeds by showing that these values are unequal and thus the graphs violate our necessary condition and thus are non-isomorphic. \par  

The algorithm begins with the decomposition of the graph adjacency matrices into a unitary and diagonal matrix, which can be done using the Xanadu StrawberryFields software package \cite{StrawF1, StrawF2}. This software contains a function that will Takagi-decompose the matrix, as well as one that will take the result of this decomposition and determine the resulting interferometer and squeezing parameters exactly. Using the latter function, which will be referred to by its name in the package, GraphEmbed, each adjacency matrix is decomposed into a unitary matrix $U$ and diagonal matrix of squeezing values $D$. With access to these values, it is now possible to calculate any subsequent order of correlation needed throughout the algorithmic process, beginning with the first. \par 
As psuedocode, where a value of $k$ is given for the mode correlation order, this can be written as:

\begin{algorithm}[H]
\caption{Calculate $k$-order Correlations}\label{alg: one}
\SetKwInOut{Input}{Input}\SetKwInOut{Output}{Output}\SetKwInOut{init}{Initialize}\SetKwInOut{calc}{Calculate}\SetKwInOut{app}{Append}
\Input{\textit{U, D}: Decomposed graph adjacency matrices, each size $M \times M$\\
$k$: Order of mode correlation}
\BlankLine
\init{$C^{k}$: empty list of size $M^{k}$}
\For{$i\leftarrow  1$ \KwTo $M$}{
    \For{$j\leftarrow 1$ \KwTo $M$}{
        \BlankLine
        \For{$k\leftarrow 1$\KwTo $M$}{
        \calc{Correlation$_{k}(i,j,...,k)$}
        \app{Correlation$_{k}(i,j,...,k)$ to $C^{k}$}
        }
    }
}
\Output{$C^{k}$: the $k$-mode correlation set for encoded graph of size $M^{k}$}
\end{algorithm}

For a pair of $M$-vertex graphs, the first order calculation results in two, $M$-element lists of correlation values. Each graph's corresponding list is sorted so they can be compared; if a value appears in one list, but not the other, they cannot be equivalent and thus the graphs cannot be isomorphic. If the sorted lists of $M$ values are equivalent, we make use of the equivalence between vertices to logically eliminate permutations from $\sigma$.\par 
Here, $\sigma$ is initialized as an $M\times M$ matrix of ones, where an element $\sigma_{i,j}=1$ is the mapping of vertex $i$ in $G_{1}$ to vertex $j$ in $G_{2}$. An element $\sigma_{i,j}=0$ when such a mapping cannot be present in any isomorphism between the graphs, such as a vertex pair with a different number of edges attached to them. The goal of the algorithm is thus to use the calculated correlation values to logically deduce that as many $\sigma_{i,j}=0$ as possible, reducing the number of possible permutations between the graphs. For example, if a unique pair of values are associate with vertex $i\in G_{1}$ and $j\in G_{2}$, the permutation matrix element $\sigma_{i,j}$ is left as one and all other elements in its row and column set to zero. A unique equivalence thus rules out $2(M-1)$ possible elements of the permutation between the $M$ vertex graphs.\par 
The situation becomes more complicated when there are multiple matching values between the lists of calculated values. If there exists an $m$-element subset in one graph's list of correlation values in which every element has the same value, and a subset of the same size with the same value exists in the other graph's list, all of the permutations that map these elements to each other must be retained. Graphs that isomorphism algorithms tend to struggle with, such as regular graphs with a high degree of symmetry, will encounter this due to the similarity of their lower-order structures.\par 
The following pseudocode is a specific example of the two-mode correlation permutation check, where a pair of $M\times M$ grids of values are compared against one another: 

\begin{algorithm}[H]
\caption{Permutation Checking}\label{alg: two}
\SetKwInOut{Input}{Input}\SetKwInOut{Output}{Output}\SetKwInOut{init}{Initialize}
\Input{$k$: Order of mode correlation\\
$C_{1}^{k}$: $k$-order correlations for $G_{1}$\\
$C_{2}^{k}$: $k$-order correlations for $G_{2}$\\
$\sigma_{k-1}$: $M\times M$ matrix of possible mappings}
\For{$i \leftarrow 1$ \KwTo $M$}{
\For{$j\leftarrow 1$ \KwTo $M$}{
\eIf{$C^{(k)}_{1}[i] = C^{(k)}_{1}[j]$}{$\sigma_{k}[i,j] = \sigma_{k-1}[i,j]\times 1$}{$\sigma_{k}[i,j] =\sigma_{k-1}[i,j]\times  0$}
    }
}
\Output{$\sigma_{k}$: Matrix of possible permutations up to order $k$}
\end{algorithm}

Once the possible mappings have been logically eliminated for the given order, the algorithm assesses if the matrix $\sigma$ can still be a valid permutation between the graphs. As the algorithm seeks to show that a permutation cannot exist between the encoded graphs, it separates $\sigma$ into one of three categories: invalid, valid, and indeterminate. An invalid permutation matrix is one with an empty row or column, which indicates there exists no mapping between the graphs that includes the corresponding vertex. This case immediately negates the necessary condition and tells the algorithm the graphs cannot be isomorphic. A valid result arises when $\sigma$ is a unique permutation matrix: it has exactly one non-zero element in each row and column and bijectively maps the vertex labels to one another. If such a matrix is found, it is then easy to verify by applying it to the graph adjacency matrices. While this result would allow the algorithm to determine isomorphism between the graphs, it is also relatively unlikely with a truncated value of correlation $k<M$.\par 
The final, and most common, categorization of $\sigma$ is as indeterminate. In this case, the matrix still has multiple non-zero elements in one or multiple rows and columns, meaning a permutation between the graphs may exist within it but it is not unique or necessarily trivial to identify. The algorithm will thus assess $\sigma$ and decide if it should manually check the graphs for isomorphism or repeat the process at higher order. The following subroutine checks if the matrix $\sigma$ can still be a valid permutation matrix at each order:

\begin{algorithm}[H]
\caption{Is Permutation Matrix}\label{alg: three}
\SetKwInOut{output}{Output}\SetKwInOut{Input}{Input}\SetKwFunction{sum}{Sum}\SetKwFunction{glynn}{Perm}\SetKwFunction{thresh}{Threshold}
\Input{$\sigma$: $M\times M$ matrix of possible permutations}
    \For{$i \leftarrow 1$ \KwTo  $M$}{
        \If{\sum{row[i]} = 0}{\output{False}}}
    \For{$i \leftarrow 1$ \KwTo  $M$}{
        \If{\sum{column[i]} = 0}{\output{False}}}
    \For{$i \leftarrow 1$  \KwTo $M$}{
        \If{\sum{row[i]} = 1}{
            \If{\sum{column[i]} = 1}{\output{True}}}}
    \eIf{\glynn{$\sigma$} $\geq$ \thresh{$k+1$}}{\output{Indeterminate, Above}}{\output{Indeterminate, Below}}
\end{algorithm}

To count the number of unique, valid permutations that exist within $\sigma$, the last step calculates the matrix Permanent, a function that is classically hard to compute exactly, even for binary matrices \cite{Valiant79}. However, it suffices to approximate the value, such as with the polynomial-time JSV algorithm for positive, real-valued matrices \cite{JSV2004}, to determine if it is below threshold. The threshold when increasing in order depends on a cost exponential in the new value of $k$ to calculate the correlation elements, as well as the cost of comparing these elements to reduce the remaining elements of $\sigma$. As the correlation order $k$ increases, the amount of values associated with each vertex grows as $M^{k-1}$ for each graph. A complete calculation of the $k$-order correlation would thus result in a $k$-dimensional structure of correlation values, such as an $M\times M$ grid for $k=2$, an $M^{3}$-element cube for $k=3$, and so on. These sets of values are checked according to the permutations remaining in $\sigma$, with progressively larger structures within the graphs being identified and compared.\par

Importantly, when we increase $k$ we bring the same permutation matrix with us from the lower order test. This follows from theorem 1, as for the encoded graphs to be isomorphic, the permutation of mode labels must map each order of correlations to one another. Therefore, if we are able to show no valid permutation matrix exists between any one of the orders of correlation it is sufficient to conclude that the graphs cannot be isomorphic. The cost of comparing the correlations also increases with order $k$, incentivizing us to eliminate as many permutations as possible at lower order to reduce the quantity of increasingly fine-grained comparisons at higher order.\par 
Often, the first-order values will not be enough to satisfy the test, necessitating an increase to second order. As permutations are applied to either rows or columns of a matrix, we are able to compare the correlations $C_{1}^{(2)}$ and $C_{2}^{(2)}$ row by row (or column by column). A list of the values in the first row of $C_{2}^{(1)}$ is compared to each row in $C_{2}^{(2)}$, with the algorithm recording when there is an equivalence between the lists of values. It is possible for a row in $C_{1}^{(2)}$ to be equivalent to multiple rows in $C_{2}^{(2)}$; in fact, it is increasingly common as the graphs become more similar. For this reason, elements $\sigma_{i,j}$ of the permutation matrix are left as one for matching rows corresponding to vertices $i$ and $j$ in the graphs, while the remaining mappings are set to zero. This process is repeated for each row in $C_{1}^{(2)}$, reducing the number of non-zero elements in $\sigma$ to refine the permutation search. \par 

Once as many elements have been eliminated from $\sigma$ as is possible with the information available at order $k$, we check if $\sigma$ is still a valid permutation matrix. If not, the order is increased to include another mode and the process repeated for the remaining elements, until IsPermutationMatrix($\sigma$) is no longer indeterminate or some truncation value of $k$ is reached. This truncation means that the algorithm will not always be able to answer the isomorphism question conclusively, especially for large, highly symmetric graphs that can only be distinguished through comparison of higher-order graph structures. \par 

With its constituent subroutines defined, the complete algorithmic process can be stated: 

\begin{algorithm}[H]
\caption{Sampler-Encoded Graph Isomorphism Test}\label{alg: main}
\SetKwInOut{Input}{Input}\SetKwInOut{Output}{Output}\SetKwInOut{init}{Initialize}\SetKwFunction{decom}{GraphEmbed}\SetKwFunction{corr}{CorrelationValue}\SetKwFunction{thresh}{Threshold}\SetKwFunction{permch}{PermCheck}\SetKwFunction{perm}{IsPermutationMatrix}
\Input{$G_{1}$: Adjacency matrix of graph 1, size $M\times M$\\
$G_{2}$: Adjacency matrix of graph 2, size $M\times M$\\
$k_{max}$: Maximum order of $k$}
$U_{1}, D_{1} \leftarrow$ \decom{$G_{1}$}\\
$U_{2}, D_{2} \leftarrow$ \decom{$G_{2}$}\\
\For{$k \leftarrow 1$ \KwTo $k_{max}$}{
    $C_{1}^{k}\leftarrow$ \corr{$U_{1}, D_{1}, k$}\\
    $C_{2}^{k}\leftarrow$ \corr{$U_{2}, D_{2}, k$}\\
    $\sigma_{k} \leftarrow$ \permch{$C_{1}^{k},C_{2}^{k},k,\sigma_{k-1}$}
    \eIf{\perm{$\sigma_{k}$}=False}{End: Graphs are not isomorphic}{\eIf{\perm{$\sigma_{k}$}=True}{End: Valid $\sigma$ found}{\eIf{\thresh = Above}{Repeat for $k+1$}{Check $\sigma_{\kappa} G_{1}\sigma^{\intercal}_{\kappa}=G_{2}\forall \sigma_{\kappa}\in \sigma_{k}$}}
    }}
\end{algorithm}

The process of calculating correlation values is polynomial in the number of vertices $M$, but scales exponentially with the order of correlation calculated $k$. As each graph has $M^{k-1}$ values calculated at each order $k$, the comparison to assess permutations can become rapidly expensive, with $M^{k^{2}}$ elements to check, if done with brute force. This motivates the logical elimination of as many permutations as possible at lower order, but the ability to do so depends strongly on the properties of the compared graphs. 

\section{Comparison with Existing Protocols}\label{sec: comp}
Our algorithm exists within a rich body of literature on the graph isomorphism problem, but does not posit paradigm-shifting advancement in computational efficiency when compared to the state-of-the-art method \cite{Babai}. Instead, our method possesses similar distinguishing power and is of comparable complexity to that which could be implemented on a physical Gaussian boson sampler \cite{GIGBS}, as well as the extensively-studied classical color refinement algorithm \cite{WL}.  

\subsection{Inspiring GBS-Based Algorithm}
Comparing our newly developed method to the inspiring process implemented on a GBS device \cite{GIGBS}, we see parallels in their overall complexity, but differences in where it arises. The number of possible output states of a sampler grows exponentially with the number of modes and photons, necessitating they be grouped together in some way to enable efficient comparison of the output distribution. The quantum method collects a large quantity of samples from the graph-encoded distributions before sorting them into orbits, equivalence classes under permutation of modes, in order to calculate the partition-averaged photon distribution \cite{GIGBS}. For each mode, the probabilities of detection events within an orbit of \textbf{n} photons, with photons in that mode, are summed together, weighted by the photon multiplicity. To coarse-grain this value, the values for different orbits are then collected to create an $M$-element list of values, one for each output mode, for each of the encoded graphs. The equivalence of these lists, up to permutation, forms the necessary condition tested for by the algorithm in \cite{GIGBS}. \par 

While the proposed method in \cite{GIGBS} efficiently collects samples from graph-encoded distributions, it shifts the complexity to the post-processing necessary in order to compare them. While both this method and ours make use of the Takagi decomposition encoding of the graphs into a sampler, we eschew collecting samples in favor of explicitly calculating statistical properties of the sampler itself. In doing so we sacrifice the efficiency of directly drawing physical samples from the distribution, but bypass the task of drawing increasingly many samples and the expensive combinatorics of preparing them for comparison. Our algorithm, for lower orders $k$, is thus of comparable complexity to the collection of samples and generation of a pair of coarse-grained distributions for comparison. 

\subsection{Classical Color Refinement Algorithm}
While our algorithm was inspired by a protocol for a quantum device, it shares a number of similarities with the well-known classical color refinement algorithm \cite{WL}. Color refinement is an iterative algorithm that assigns labels, in the form of colors, to vertices in a graph based on the properties of their neighborhood. Initially, each vertex is assigned the same color. Each pair of vertices are then asked if they are the same color, and if they have the same number of vertices in their respective neighborhoods; if not, they are given a different color label. In the next step, if a pair of vertices are still the same color, they are asked if the vertices in their neighborhood are the same, and again are given a different color if not. This process repeats until every pair of vertices are different colors, or those that are the same color have identically colored neighborhoods as well. Once the process has finished, the graphs can be determined to be non-isomorphic if a bijective mapping between the set of labels can be shown not to exist. The existence of this mapping is, like those between our correlators, not sufficient to conclude the presence of isomorphism, but necessary and thus able to identify non-isomorphism. \par 
Color refinement is in fact a specific version of the Weisfeiler-Leman heuristic test \cite{WL}, which generalizes the process to multiple orders. This order $k$ determines the size of vertex-tuples the above iterative process is applied to, and aligns with our interpretation of order as the size of subgraphs (number of modes/vertices) being compared. The aforementioned first-order WL test is comparable to our second-order mode correlation in complexity and distinguishing power. Generally, both methods scale exponentially with order $k$ and polynomially in the number of vertices $M$. Both methods struggle with regular and symmetric graphs, as they limit the ability to identify unique substructures. \par 
Notably, it is known that for an order $k$ of WL test performed, it is always possible to define a pair of graphs such that the test will fail to distinguish them \cite{WalksGI, CFI}. This result was proven by Cai, F\"{u}rer, and Immerman \cite{CFI, CFI2} and states more formally that there is no fixed $k$ such that a $k$-order WL test is a complete test for isomorphism. It is an open question if this result extends to our algorithm, or if there are any specific graph families that would be provably unsolvable with our method. While both our method and color refinement attempt to determine non-isomorphism by reducing the space of permutations to none, the information they use about the graphs to do so differs. The recovery of a sufficient condition at maximal order, if only theoretically computable, is another unique property that differentiates our method. For these reasons, we believe our method to be distinct from the Weisfeiler-Leman test for graph isomorphism.

\section{Conclusion}
In our study of the feasibility of realizing quantum advantage by solving the graph isomorphism problem with Gaussian boson sampling, we have developed a classical algorithm inspired by the properties of a quantum device. The existence of a sufficient condition for sampler-encoded graph isomorphism allowed us to extrapolate our own necessary condition, making use of the equivalence of the associated probability distributions. This condition can be tested with statistical quantities we are able to explicitly calculate after decomposing a graph adjacency matrix, returning us to the classical realm. We see that this approach resembles the necessary condition testable with a Gaussian boson sampler, as both methods accumulate marginal components of the encoded-graph distribution in hope of showing they cannot be equivalent. Neighbored on the quantum side by the sampling algorithm, our algorithm also finds itself adjacent to the classical Weisfeiler-Leman method in terms of complexity and distinguishing power. Our method, as well as the other two mentioned, are categorically less efficient than Babai's state-of-the-art classical method.  It remains an open question if there are specific families or types of graphs that are able to be distinguished with either greater efficiency or accuracy using our method than those in literature. This collection of necessary condition-negating heuristics agree that there remains no generally efficient quantum algorithm to solve graph isomorphism as means to demonstrate computational advantage with a near-term linear optical quantum computer. \par  

\section{Acknowledgments}
We thank G. Rattan for insight on the Weisfeiler-Leman method and the current state of isomorphism research, as well as F.H.B Somhorst for scientific discussions. We also thank Y. Cardin and N. Quesada for making us aware of \cite{Cardin2024}, the results of which provided further graphical insights into our own work. This publication is part of the project ``At the Quantum Edge” of the research programme VIDI, which is financed by the Dutch Research Council (NWO). S. N. van den Hoven acknowledges support from the PhotonDelta National Growth Fund program. 
\bibliographystyle{unsrt}
\bibliography{refs}

\end{multicols}
\end{document}